\documentclass[11pt]{article}
\begin{document}
\centerline {\large \bf The Sznajd model of consensus building with limited 
persuasion} 

\bigskip
\centerline{Dietrich Stauffer}

\centerline{Institute for Theoretical Physics, Cologne University}

\centerline{D-50923 K\"oln, Euroland}

\bigskip
{\small The Sznajd model, where two people having the same opinion can convince 
their neighbours on the square lattice, is modified in the sense of 
Deffuant et al and Hegselmann, that only neighbours of similar opinions can be 
convinced. Then consensus is easy for the competition of up to three opinions 
but difficult for four and more opinions.}

Keywords: Sociophysics, Monte Carlo, square lattice 
\bigskip

Studies of Ising models and cellular automata for social phenomena have a 
long tradition \cite{old}.
The Sznajd model \cite{sznajd} is one of the current sociophysics models of
human interactions.
It assumes that a pair of neighbouring sites (``people'') on a square or other 
lattice convince their neighbours if and only if the two people in the pair 
share the same opinion: ``United we stand, divided we fall''. Bernardes 
already generalized it from two to $q$ opinions, to describe election results
\cite{bernardes}. One finds that starting with a random initial distribution
of opinions, after some time a consensus is formed when everybody has the
same opinion. 

From a different direction, Deffuant et al \cite{deffuant} and Hegselmann
\cite{hegselmann}
simulated persuasion processes with a continuous spectrum of opinions between
0 and 1, where 
one person can convince neighbours of the own central opinion only if their
opinion differ by a small amount $\epsilon$ from the central opinion. 
Replacing $1/\epsilon$ by $q$ we get a similar discrete model with opinions 
1,2, ... $q$, if only neighbours differing in their opinion by $\pm 1$ from 
the central opinion are convinced. We now combine this model with the Sznajd
principle that only pairs of identical opinions are convincing. 

Thus every ``Potts'' spin $S_i = 1,2, \dots, q$ on a square lattice is in one
of $q$ different states. A randomly selected pair $<S_i, S_k>$ of nearest 
neigbours $<i,k>$ may convince its six neighbours only if $S_i=S_k$; 
then it actually convinces a neighbour $j$ if and only if $|S_j-S_i| = 1$. 

It is plausible and well known from building coalitions in parliaments that
consensus is the more difficult the larger the number $q$ of opinions is.
For $q=2$ in the present model, each opinion is close enough to the other 
opinion to be convinced, the restriction  $|S_j-S_i| \le 1$ becomes meaningless,
and the standard Sznajd model is recovered. For $q=3$ at least opinion 2 can
convince everybody with a different opinion. For $q=4$, on the other hand,
opinion 1 is well separated from opinion 4, and a phase separation into 
different never changing domains is possible. We check by Monte Carlo simulation
whether such theoretically possible results actually are reached.

Thus we start with the initial opinions randomly distributed with equal weight 
over the square lattice, and follow then the above kinetics until all spins
are the same, or until nothing changes anymore, for $31 \times 31$ to
$301 \times 301$ lattices. We then found that for $q=3$ nearly always all
spins became parallel, $S_i=2$, after some time, while for $q=4$ nearly always 
they got stuck in an inhomogeneous fixed point. In rare cases, $q=3$ went to an
inhomogeneous fixed configuration, and $q=4$ to a never-ending dynamics or to
all parallel.

For $q=3$ the dominance of the centrist opinion $S_i=2$ over the two more
extreme opinions 1 and 3 comes from the fact that 1 and 3 only can convince
neighbours of opinion 2, while opinion 2 can change opinions 1 and 3. If instead
we assume cyclicity, that opinion 1 can convince both 2 and $q$, and opinion 
$q$ can convince both opinion 1 and $q-1$, then all $q$ states become 
equivalent. Actually, for $q=3$ in this case one has no limit on persuasion 
since each opinion can change the other two opinions, and we return to the 
standard Sznajd model with $q$ opinions \cite{bernardes} where for all $q$ at
the end all spins seem to be parallel. For $q \ge 4$, there is still a limit
on persuasion, and again we end up mostly in inhomogeneous fixed points. Thus 
the cyclicity does not change the boundary between $q=3$ and 4: Usually, for 
$q \le 3$ we reach consensus and for $q \ge 4$ we don't. 

It depends on the interpretation of the model whether we regard the case of
all spins parallel as desirable (consensus) or not (dictatorship). 
  
I thank the SimSoc5 conference \cite{hegselmann} for the invitation to 
participate, and R. Hegselmann for literature information.


\begin{thebibliography}{99} 
\bibitem{old}J.M. Sakoda, J. Math. Sociol. 1, 119 (1971); T. Schelling, J. Math.
Sociol. 1, 143 (1971); E. Callen and D. Shapero, Phys. Today July 1974, p. 23;
W. Weidlich: Sociodynamics; A Systematic Approach to
Mathematical Modelling in the Social Sciences. Harwood Academic Publishers, 
2000; for an example of recent research see A. Flache and R. Hegselmann,
Journal of  Artificial Societies and Social Simulation 4, No. 4 (2001)
(electronic only: http://www.soc.surrey.ac.uk/JASSS/4/4/6.html)

\bibitem{sznajd} K. Sznajd-Weron and J. Sznajd, Int. J. Mod. Phys. C 11,
1157 (2000); D. Stauffer, A.O. Sousa and S. Moss de Oliveira, Int. J. Mod.
Phys. C  11, 1239 (2000); K. Sznajd-Weron and R. Weron, Int. J. Mod. Phys. C 13,
No. 1 (2002); A.A. Moreira, J.S. Andrade Jr. and D. Stauffer, Int. J. Mod.
Phys. C  12, 39 (2001); Ren\'e Ochrombel, Int. J. Mod. Phys. C 12, 1091 (2001);
I. Chang, Int. J. Mod. Phys. C 12, No. 10 (2001); A.S. Elgazzar, Int. J. Mod. 
Phys. C 12, No. 10 (2001); J. Schneider and S. Moss de Oliveira, priv. comm.

\bibitem{bernardes} A.T. Bernardes, U.M.S. Costa, A.D. Araujo, and D. Stauffer, 
 Int. J. Mod. Phys. C  12, 159 (2001);  A.T. Bernardes, D. Stauffer and J. 
Kert\'esz, Eur. Phys. J. B preprint.

\bibitem{deffuant} G. Deffuant, D. Neau, F. Amblard and G. Weisbuch, 
Adv. Complex Syst. 3, 87 (2000)

\bibitem{hegselmann} R. Hegselmann, talk at ``Simulating Society V: Frontiers
in Social Sciences Simulations'', Kazimierz Dolny, September 2001, organized
by A. Nowak, to be published as ``Opinion dynamics under bounded confidence: 
Models and simulations'' by  R. Hegselmann and M. Krause in Journal of  
Artificial Societies and Social Simulation (conference proceedings). See also
J.C. Dittner, Nonlinear Analysis 47, 4615 (2001).
\end{thebibliography}
\end{document}